\documentclass{XrU2005}
\usepackage{epsfig}

\title{Obscured AGN: clues from high-resolution imaging and spectroscopy}
\author{Stefano Bianchi}
\author{Matteo Guainazzi}
\affil{XMM-Newton Science Operations Center, ESAC, ESA, Apartado 50727, E-28080 Madrid, Spain}
\author{Marco Chiaberge}
\affil{Space Telescope Science Institute, 3700 San Martin Drive, Baltimore, MD 21218}

\begin{document}


\maketitle

\begin{abstract}
We present a sample of 8 Seyfert 2 galaxies observed by \textit{HST}, \textit{Chandra} and XMM-\textit{Newton}. All of the sources present soft X-ray emission which is coincident in extension and overall morphology with the [{O\,\textsc{iii}}] emission. The spectral analysis reveals that the soft X-ray emission of all the objects is likely to be dominated by a photoionized gas. We tested with the code \textsc{cloudy} a simple scenario where the same gas photoionized by the nuclear continuum produces both the soft X-ray and the [{O\,\textsc{iii}}] emission. Solutions satisfying the observed ratio between the two components exist, and require the density to decrease with radius roughly like $r^{-2}$, similarly to what often found for the Narrow Line Region.
\end{abstract}

\section{The sample}

The sample consists of all the Seyfert 2 galaxies included in the \citet{schm03} catalog, with a \textit{Chandra} observation, with the only exclusion of NGC~1068, which has already been extensively studied. On the other hand, we added another source, NGC~5643. All the sources have also an XMM-\textit{Newton} RGS observation, except for NGC~5347.

\section{Analysis}

\subsection{Imaging}

\begin{figure*}

\begin{center}
\epsfig{file=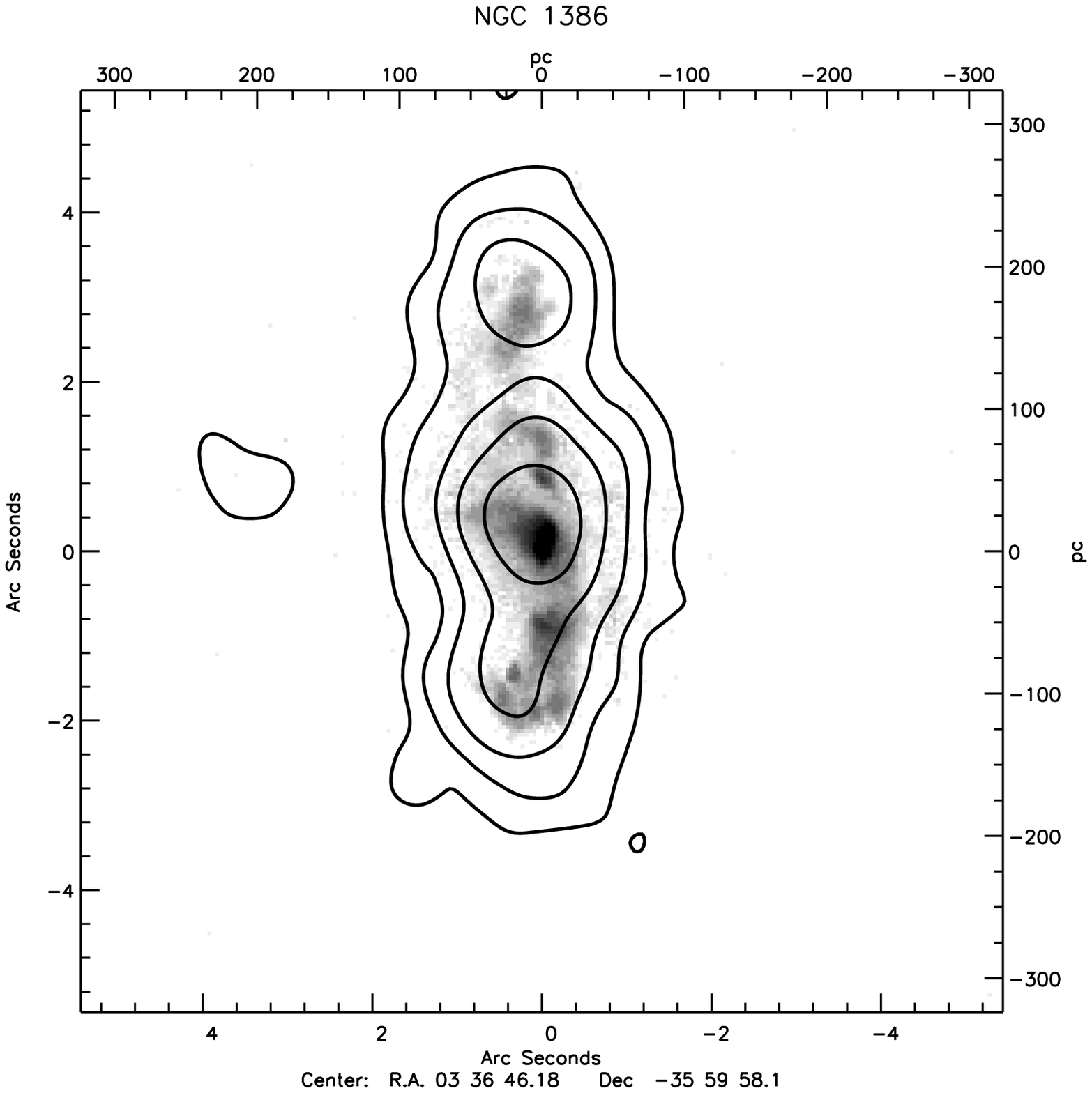, width=6.92cm}
\hspace{0.5cm}
\epsfig{file=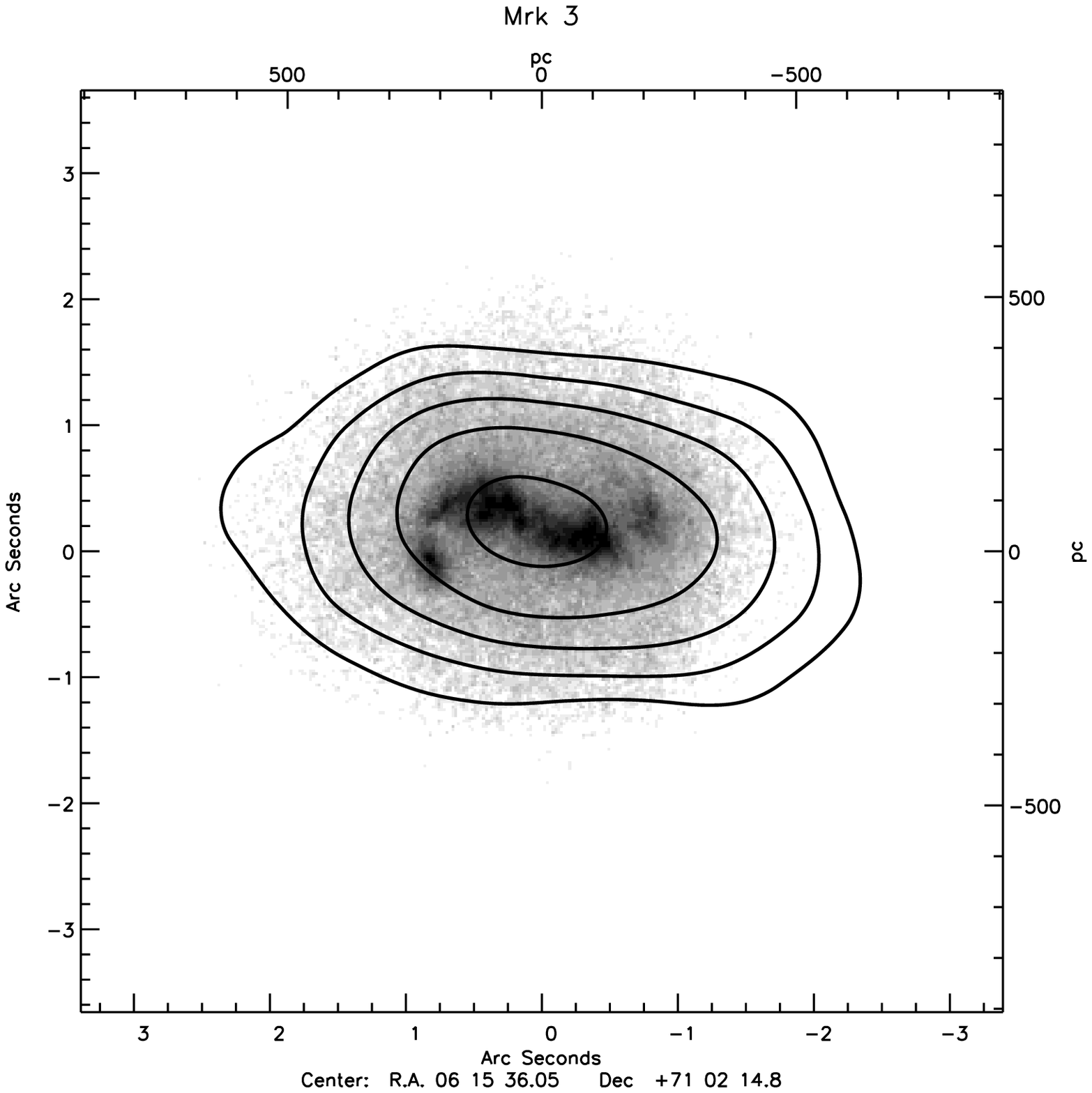, width=6.92cm}

\vspace{0.5cm}

\epsfig{file=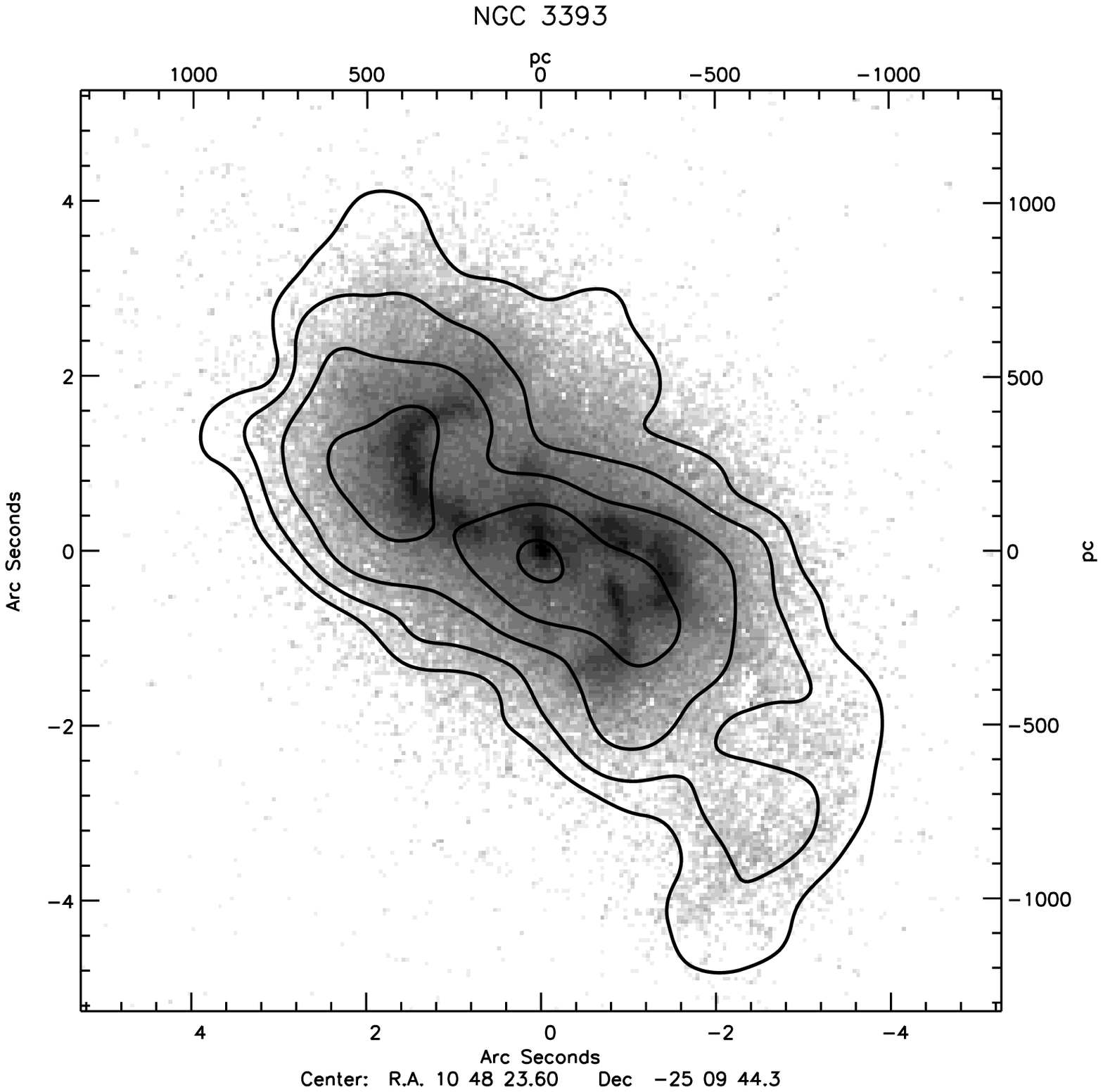, width=6.92cm}
\hspace{0.5cm}
\epsfig{file=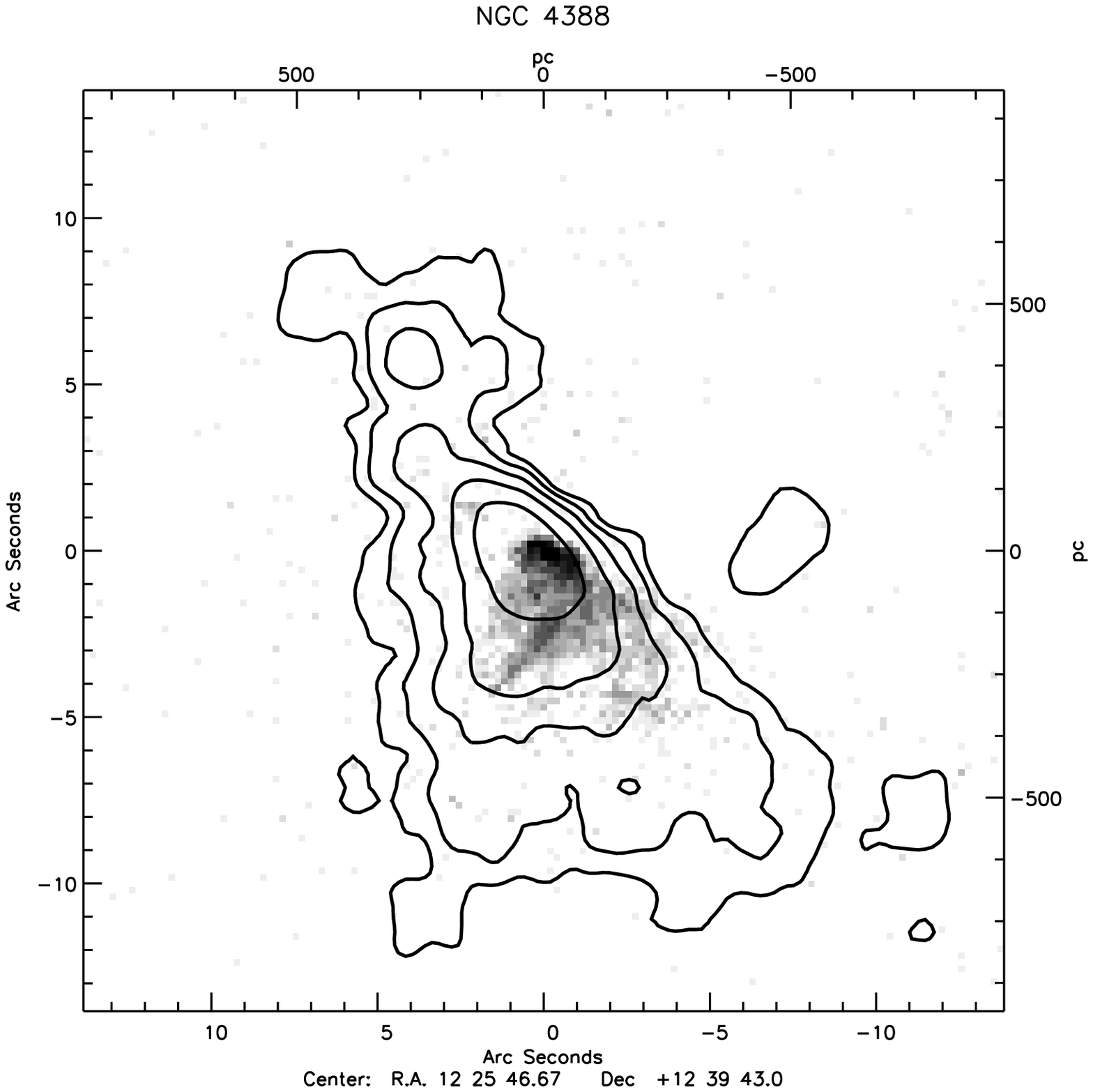, width=6.92cm}

\vspace{0.5cm}

\epsfig{file=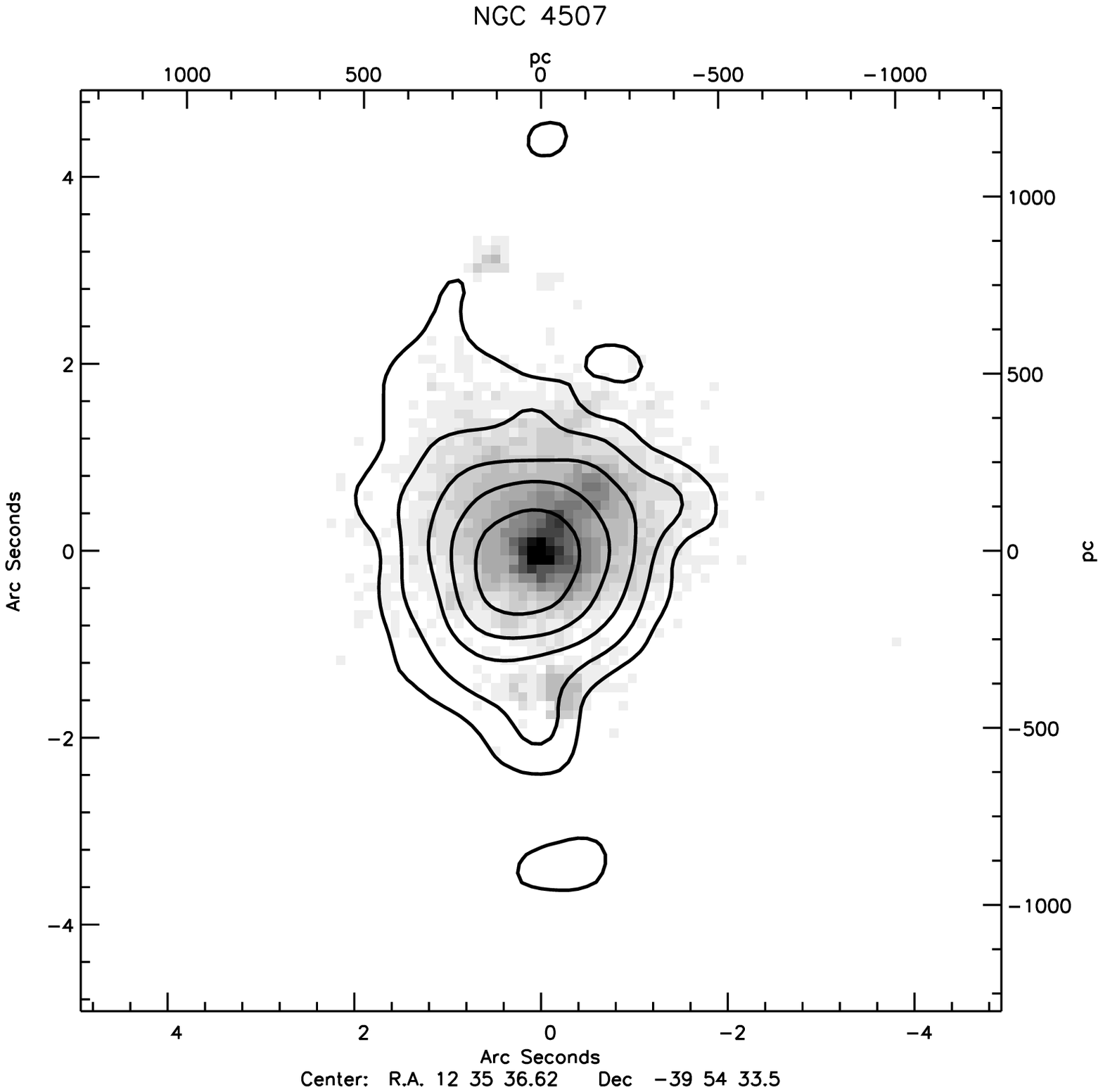, width=6.92cm}
\hspace{0.5cm}
\epsfig{file=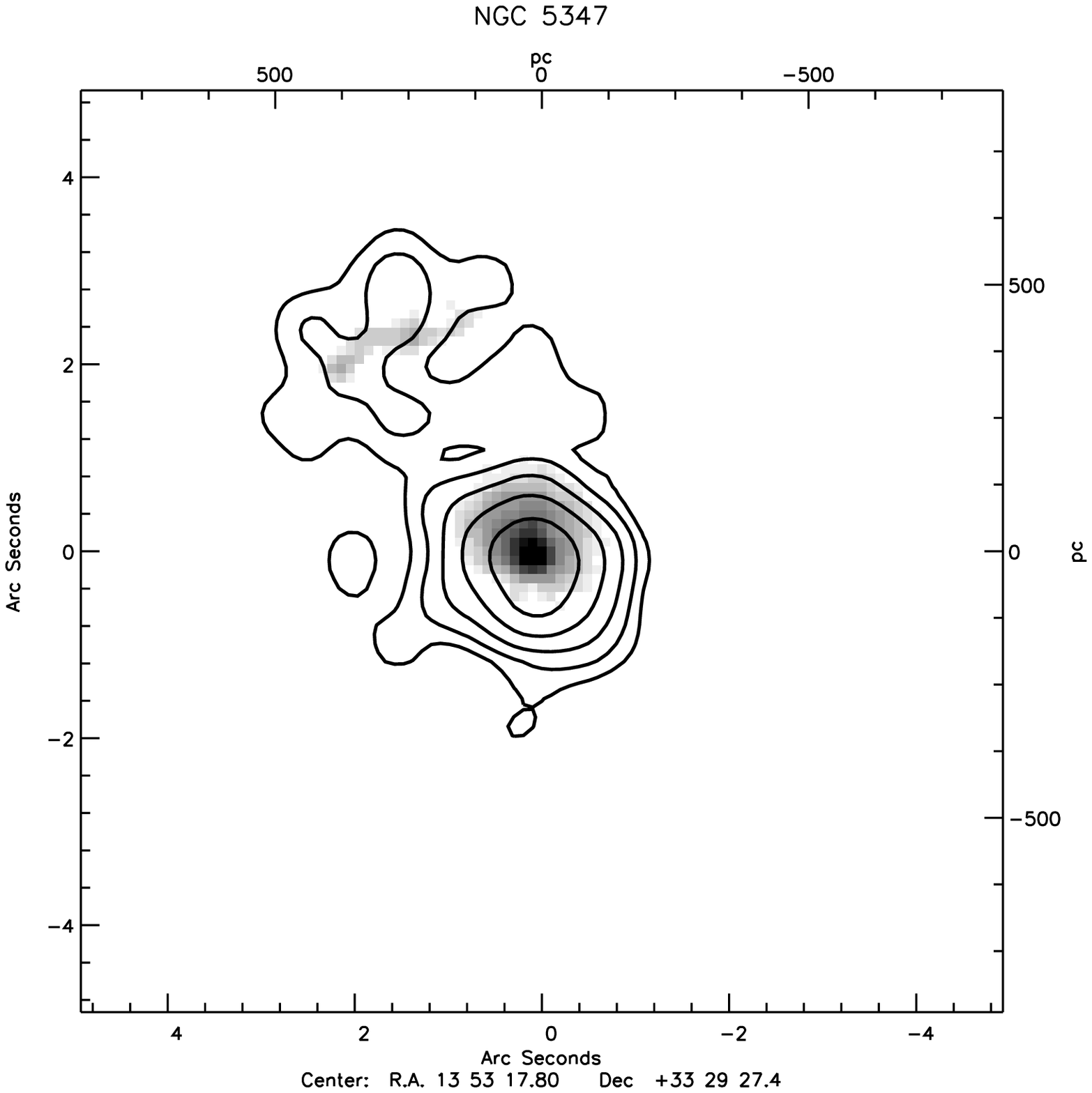, width=6.92cm}

\end{center}

\caption{\label{xray2oiiimaps}\textit{Chandra} soft X-ray contours superimposed on \textit{HST} [{O\,\textsc{iii}}] images. The contours correspond to five logarithmic intervals in the range of 1.5-50\% (NGC~1386), 5-80\% (Mrk~3), 5-90\% (NGC~3393), 4-50\% (NGC~4388), 0.5-50\% (NGC~4507) and 2-50\% (NGC~5347) of the peak flux. The \textit{HST} images are scaled with the same criterion for each source, with the exception of NGC~4388 and NGC~4507, whose [{O\,\textsc{iii}}] emission goes down to the 2\% and the 0.1\% of the peak, respectively.}
\end{figure*}

Figure \ref{xray2oiiimaps} shows the contours of the \textit{Chandra} soft X-ray emission superimposed on the \textit{HST} [{O\,\textsc{iii}}] images, for all the sources in our sample. The coincidence between the soft X-ray and [{O\,\textsc{iii}}] emission is striking, both in the extension and in the overall morphology. Unfortunately, the lower spatial resolution of \textit{Chandra} with respect to \textit{HST} does not allow us to perform a detailed comparison of the substructures apparent in the latter. 

\begin{figure*}

\begin{center}

\epsfig{file=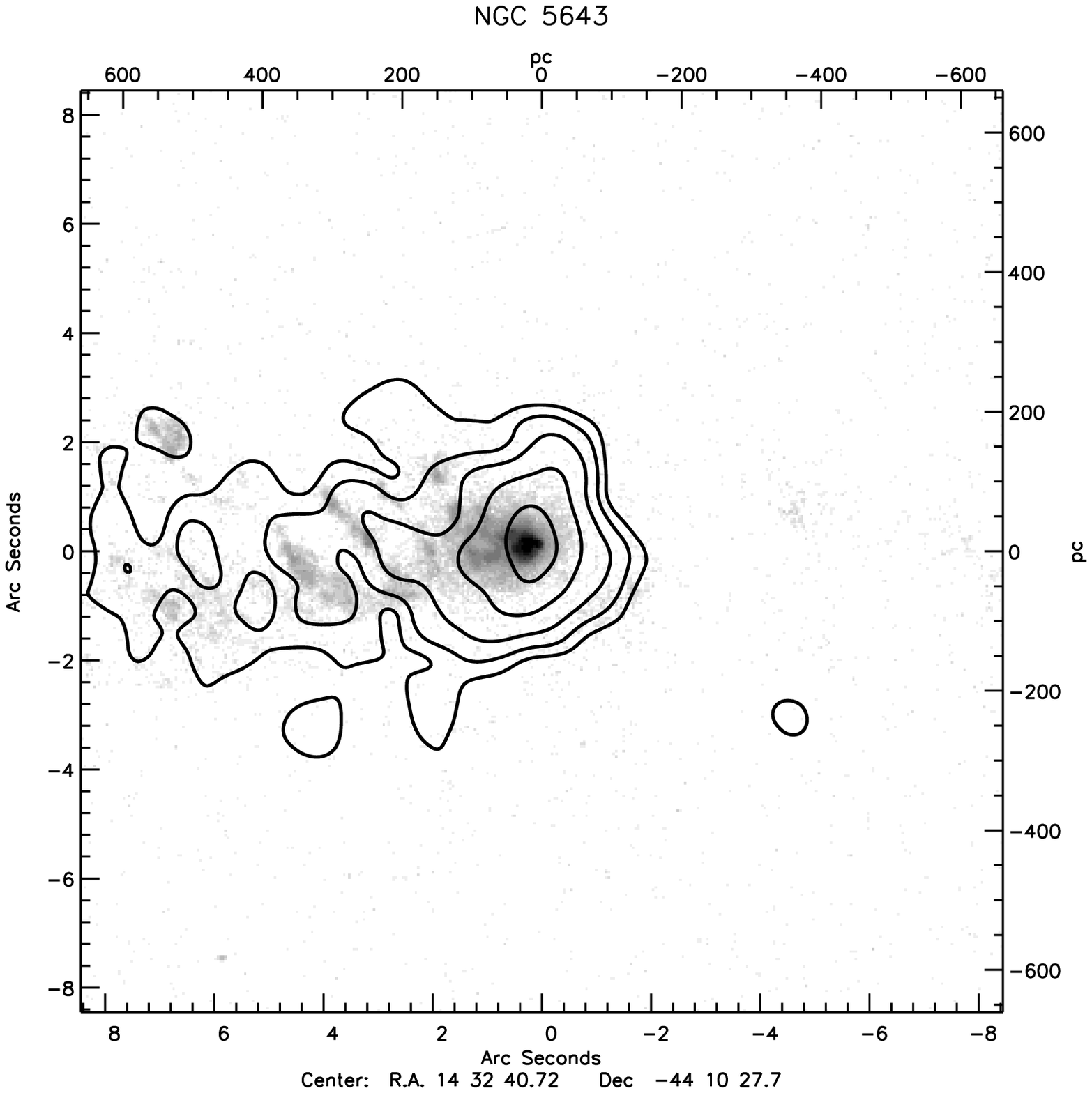, width=6.92cm}
\hspace{0.5cm}
\epsfig{file=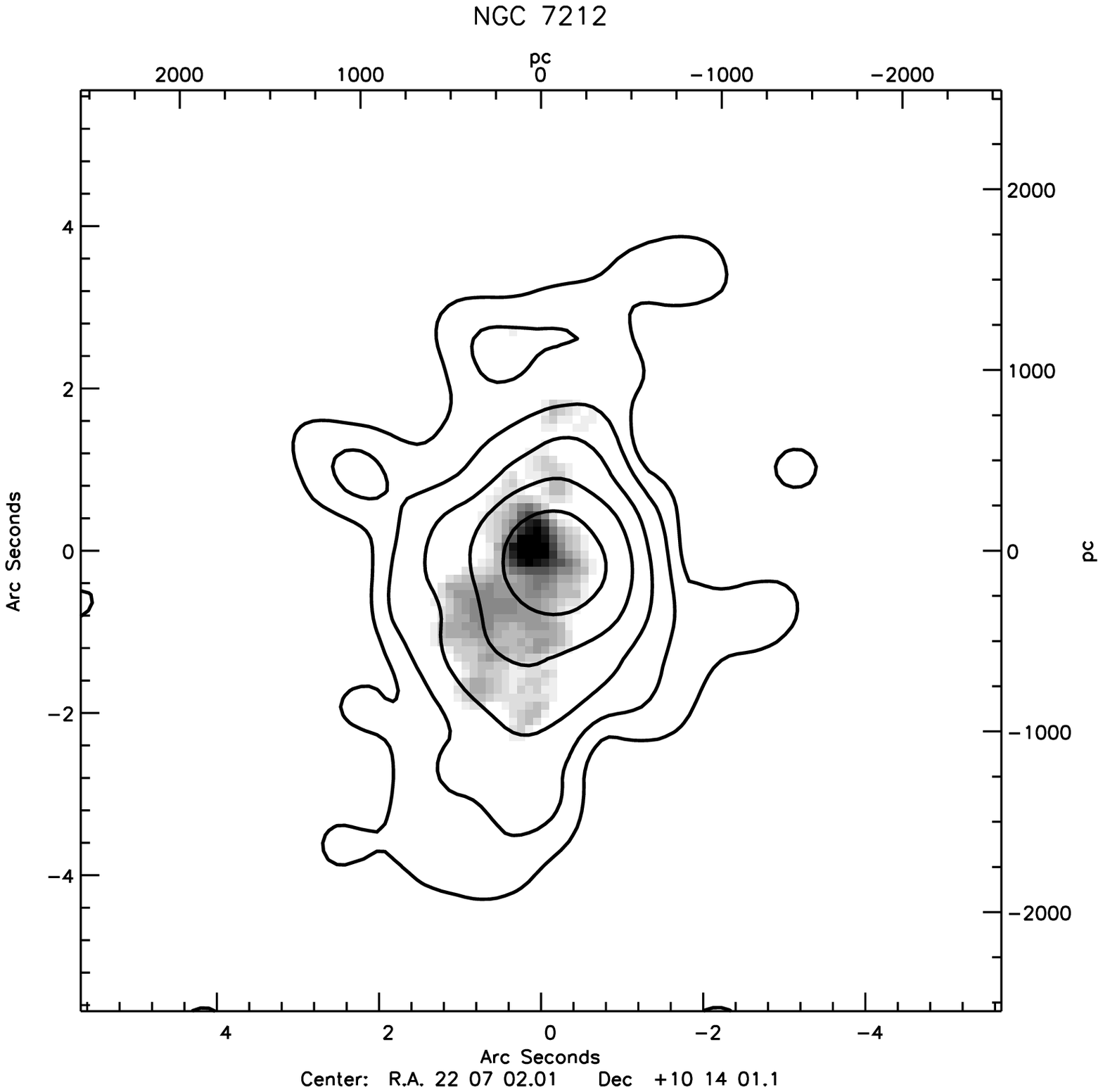, width=6.92cm}

\end{center}

\caption{\label{xray2oiiimaps_2}Same of Fig. \ref{xray2oiiimaps}, but for NGC~5643 (8-80\%) and NGC~7212 (1-50\%). In the case of NGC~5643, the [{O\,\textsc{iii}}] emission goes down to the 0.5\% of the peak.}
\end{figure*}

\subsection{\label{spectral}Spectral analysis}

\begin{figure*}

\begin{center}
\epsfig{file=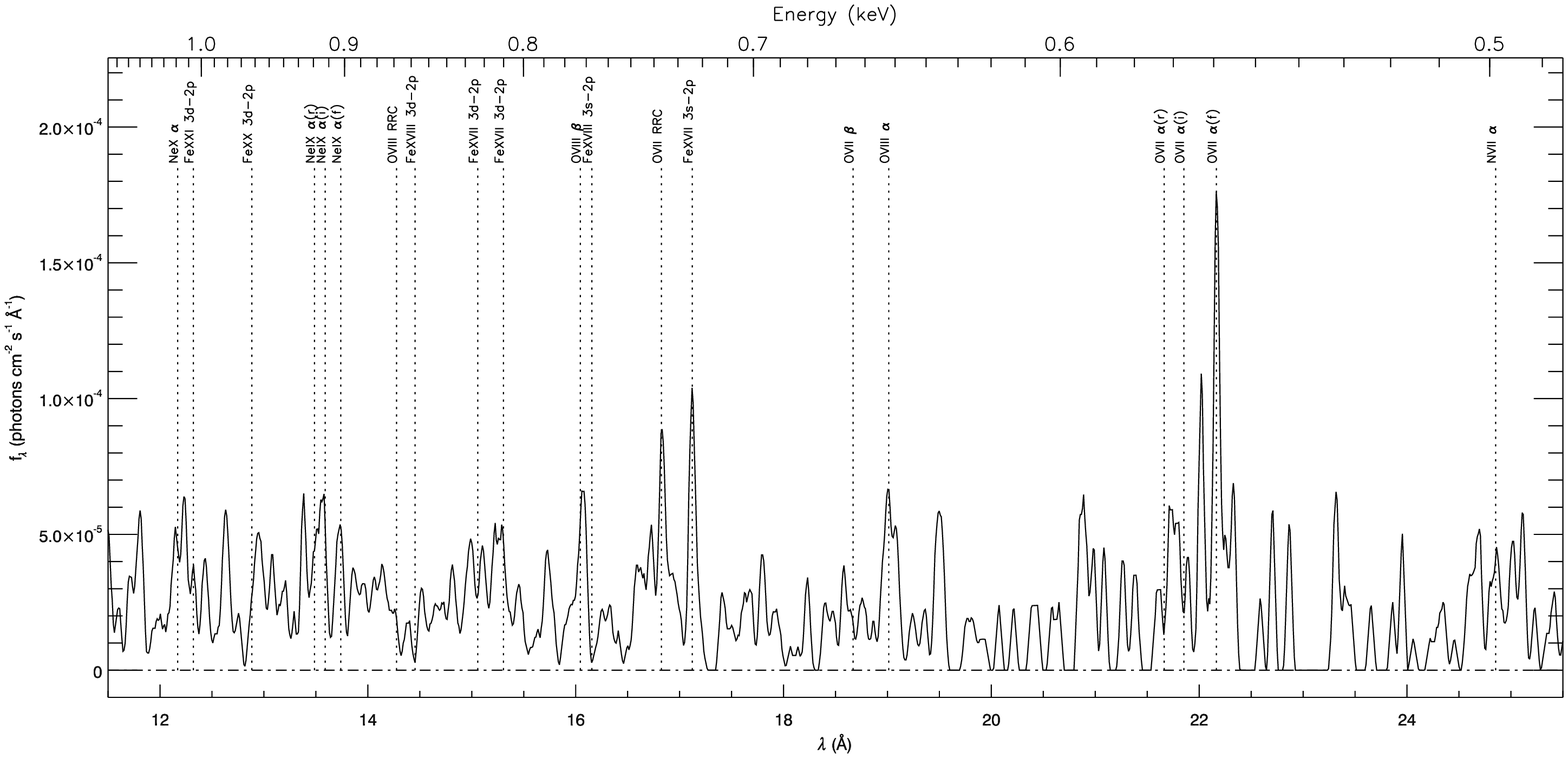, width=17.2cm,height=7cm}
\epsfig{file=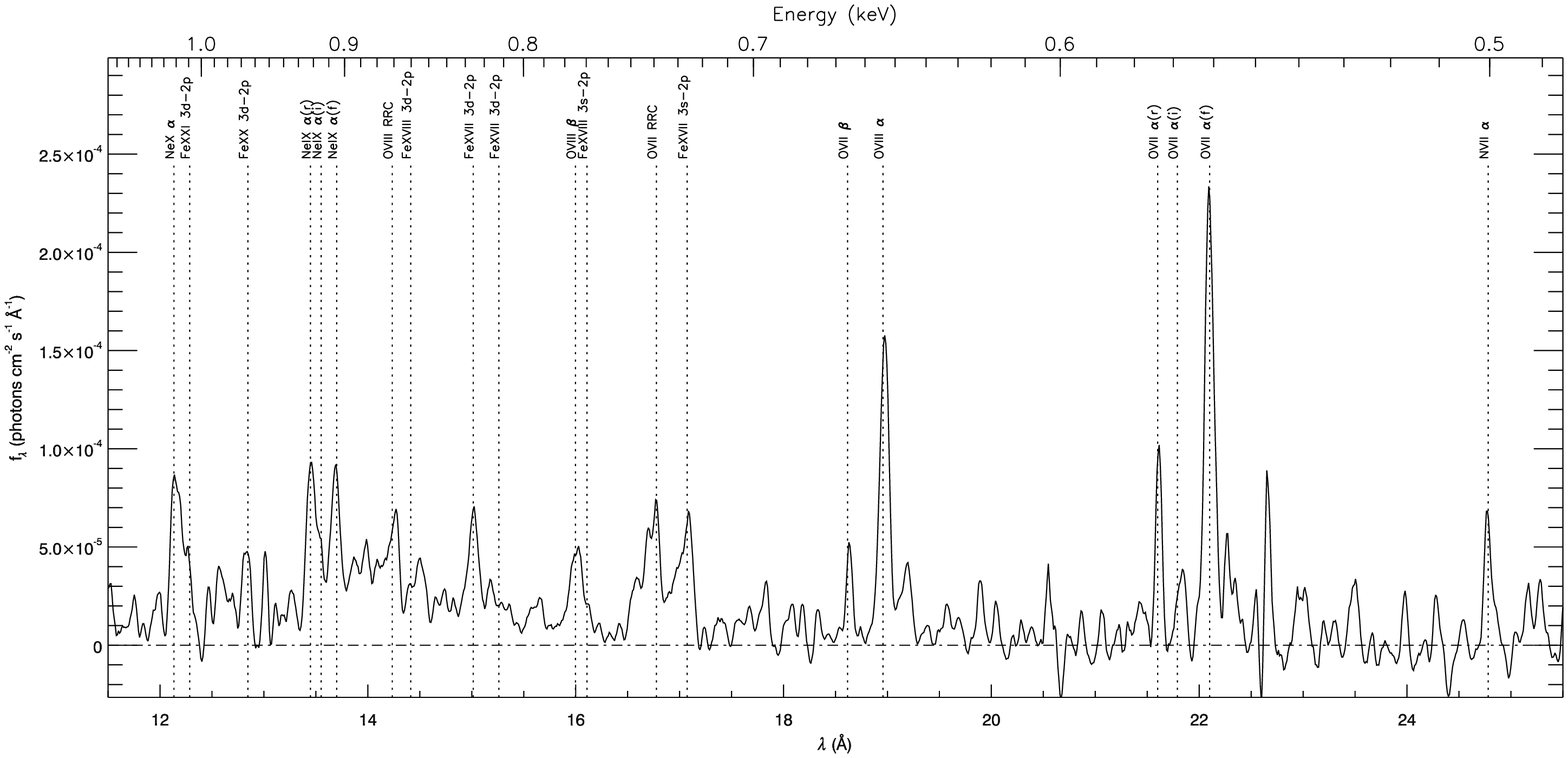, width=17.2cm,height=7cm}
\epsfig{file=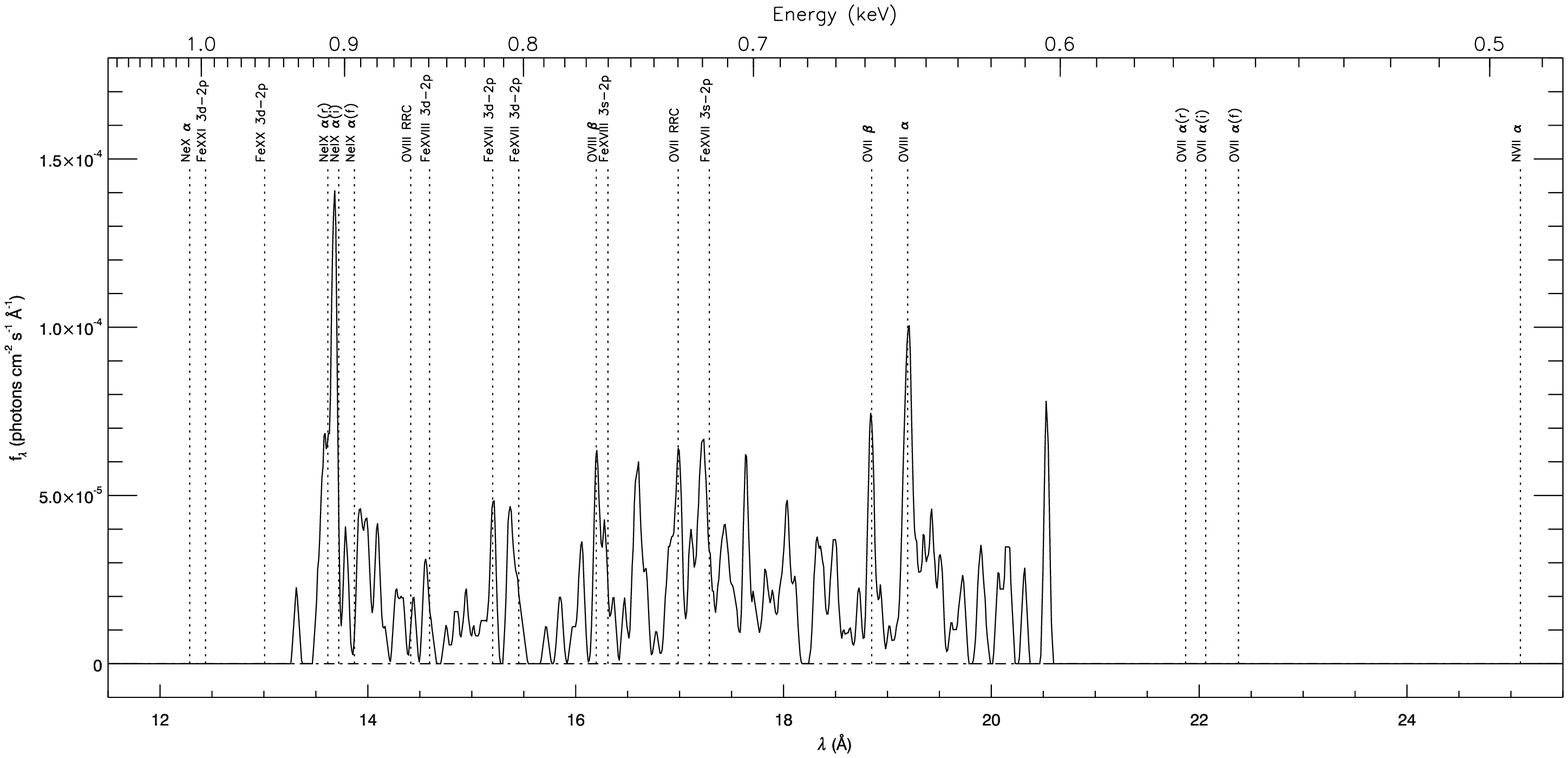, width=17.2cm,height=7cm}
\end{center}

\caption{\label{rgs1} Combined RGS1/RGS2 spectrum plotted in the rest frame of the source (12-25 $\AA$), for NGC~1386 (17 ks, F$_\mathrm{0.5-2\,keV}=1.8\times10^{-13}$ cgs), Mrk~3 (190 ks, 4.7) and NGC~3393 (14 ks, 2.2).}
\end{figure*}

\begin{figure*}

\begin{center}
\epsfig{file=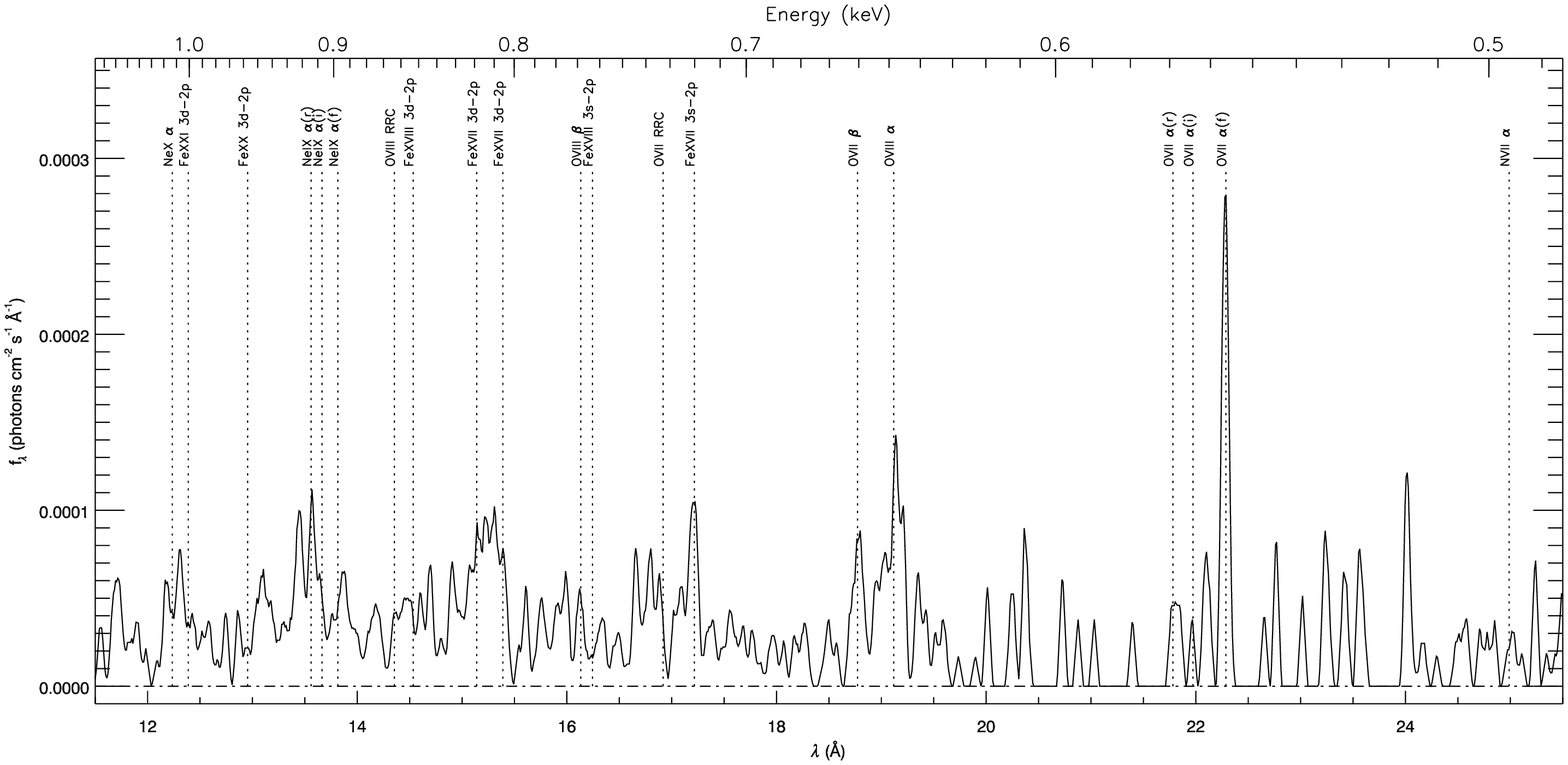, width=17.2cm,height=7cm}
\epsfig{file=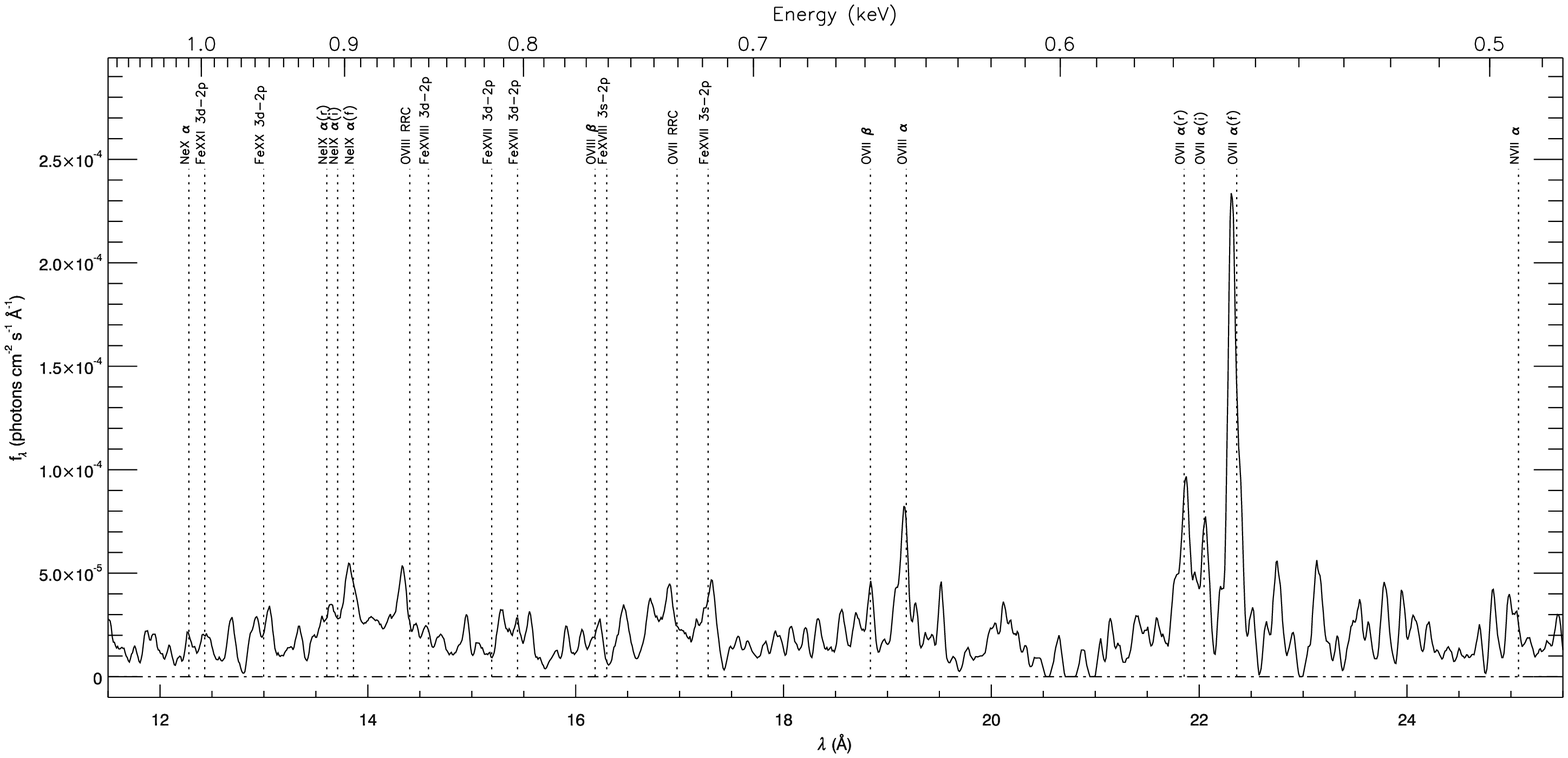, width=17.2cm,height=7cm}
\epsfig{file=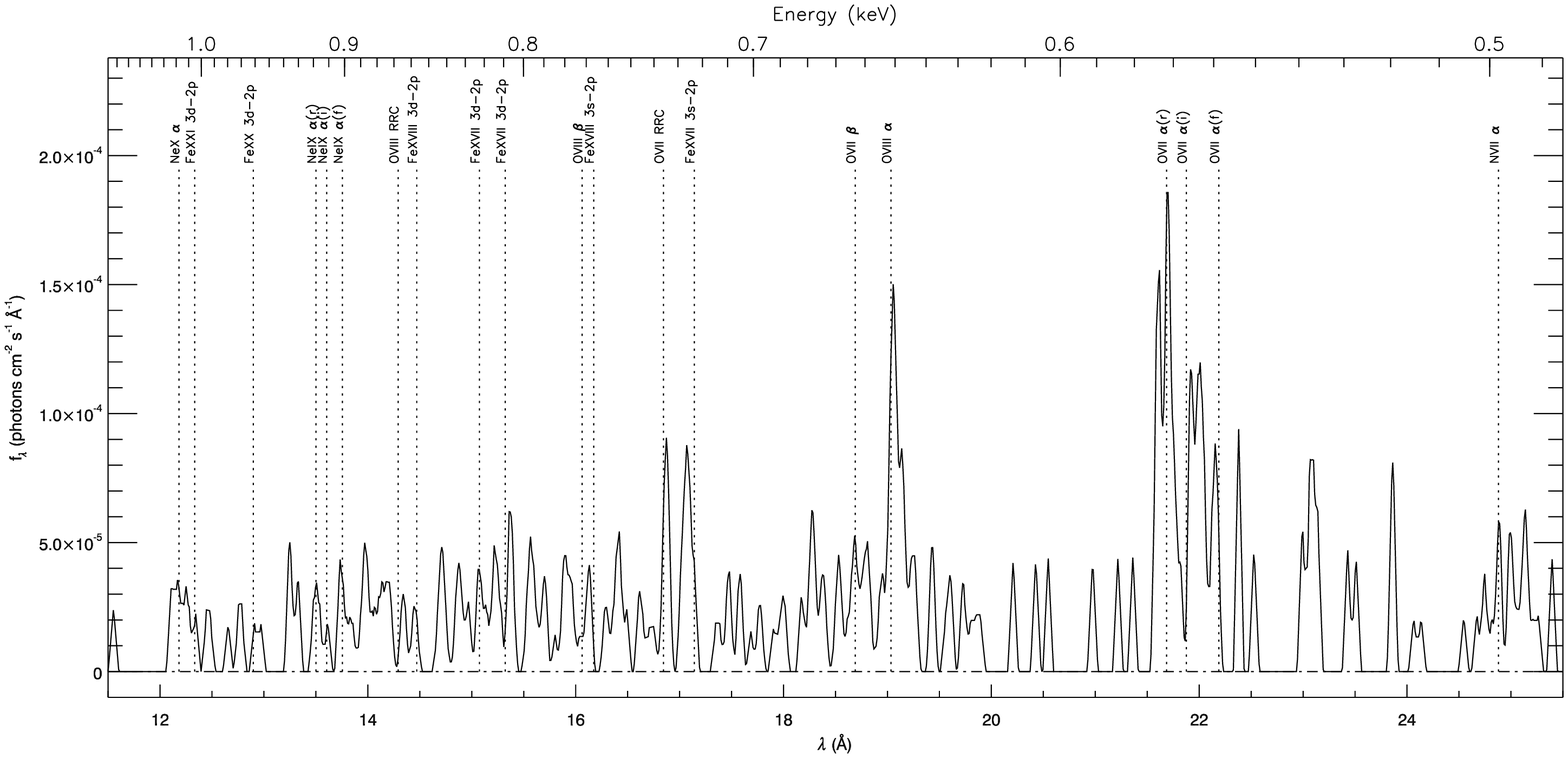, width=17.2cm,height=7cm}
\end{center}

\caption{\label{rgs2}Same as Fig. \ref{rgs1}, for NGC~4388 (12 ks, 3.4), NGC~4507 (46 ks, 3.3) and NGC~5643 (10 ks, 1.4).}
\end{figure*}

\begin{figure*}

\begin{center}
\epsfig{file=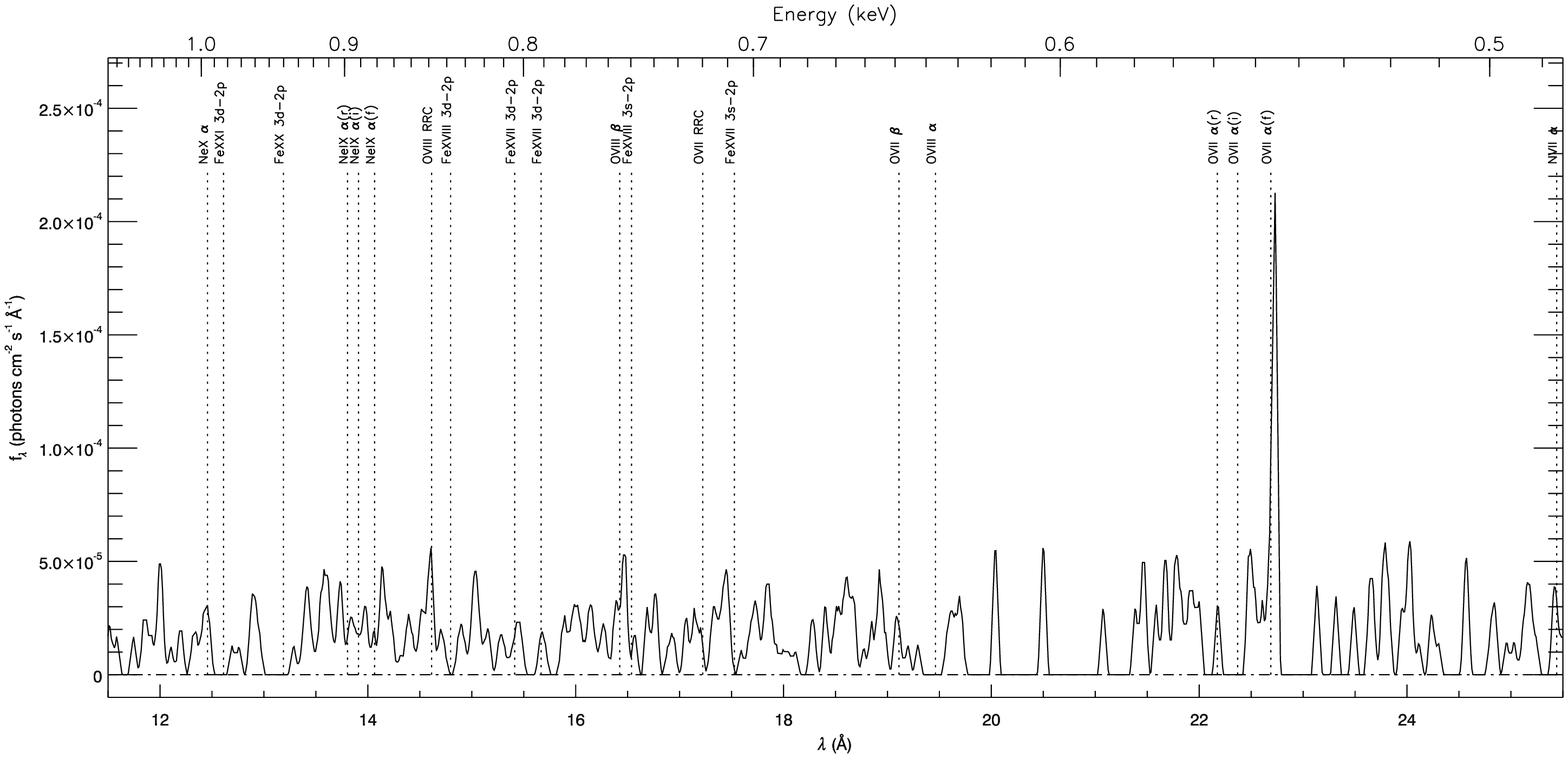, width=17.2cm,height=7cm}
\end{center}

\caption{\label{rgs3}Same as Fig. \ref{rgs1}, for NGC~7212 (14 ks, 0.8).}
\end{figure*}

The spectral analysis of the sources suggests that the most likely origin for the soft X-ray emission is in a gas photoionized by the nuclear continuum. In spectra with CCD resolution, a `scattering' model (a powerlaw plus emission lines) is to be preferred to a `thermal' model either on statistical grounds or because of unphysical best fit parameters of the latter (quasi-zero abundances). On the other hand, the RGS spectra are clearly dominated by emission lines, with a very low level of continuum (Fig. \ref{rgs1}, \ref{rgs2}, \ref{rgs3}). This allows us to easily detect strong emission lines even in short observations of objects with relatively low fluxes. The clearest piece of evidence comes from the 190 ks combined RGS spectrum of Mrk~3, which is produced in a photoionized gas with an important contribution from resonant scattering. 

The other spectra do not have enough statistics to allow us to draw unambiguous conclusions on any individual object. The predominance of the forbidden transition of the {O\,\textsc{vii}} triplet is a common feature of all spectra, except for NGC~5643. Deeper high-resolution observations of our sample will be able to confirm whether photoionization by the AGN is indeed the dominant mechanism responsible for the soft X-ray emission in these objects.

\section{Photoionization models}

The spectral analysis of the sources in our sample suggests that their soft X-ray spectra are likely dominated by emission lines produced in a material photoionized by the central AGN. On the other hand, the striking resemblance of [{O\,\textsc{iii}}] structures with the soft X-ray emission favors a common origin for both components. Therefore, since the NLR is generally believed to be also produced mainly by photoionization, we generated a number of models in order to investigate if a solution in terms of a single photoionized material to produce the optical NLR and the soft X-ray emission is tenable.
Calculations were performed with version 96.01 of \textsc{Cloudy}, last described by \citet{cloudy}. The adopted model is represented by a conical geometry, which extends from 1 to 350 pc from the nucleus, with temperature set by photoionization equilibrium under a typical AGN continuum \citep{korista97}. We produced a detailed grid of models, as a function of the following parameters: ionization parameter \textit{U} of the illuminated face of the gas; filling factor of the gas \textit{f}; density of the gas $n_{e}$. We assume a power-law radial dependence of the density, $n_e(r_0)\, \left( \frac{r}{r_0} \right) ^{-\beta}$, with $\beta=0$ (constant density) and varying between 1 and 2.4. We limit our models to a minimum total column density of $10^{20}$ cm$^{-2}$. 

\begin{figure*}
\begin{center}
\epsfig{file=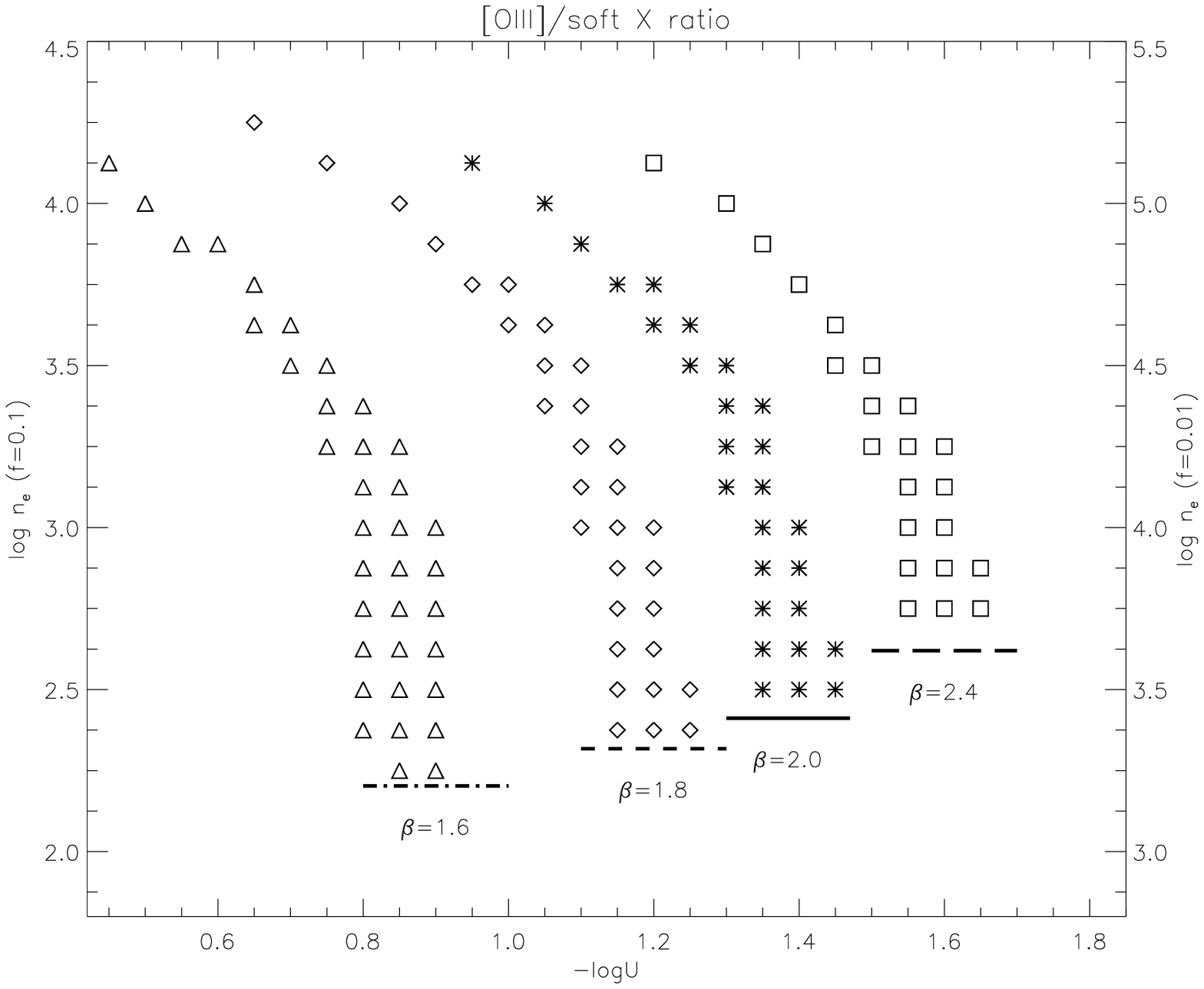, width=8cm}
\epsfig{file=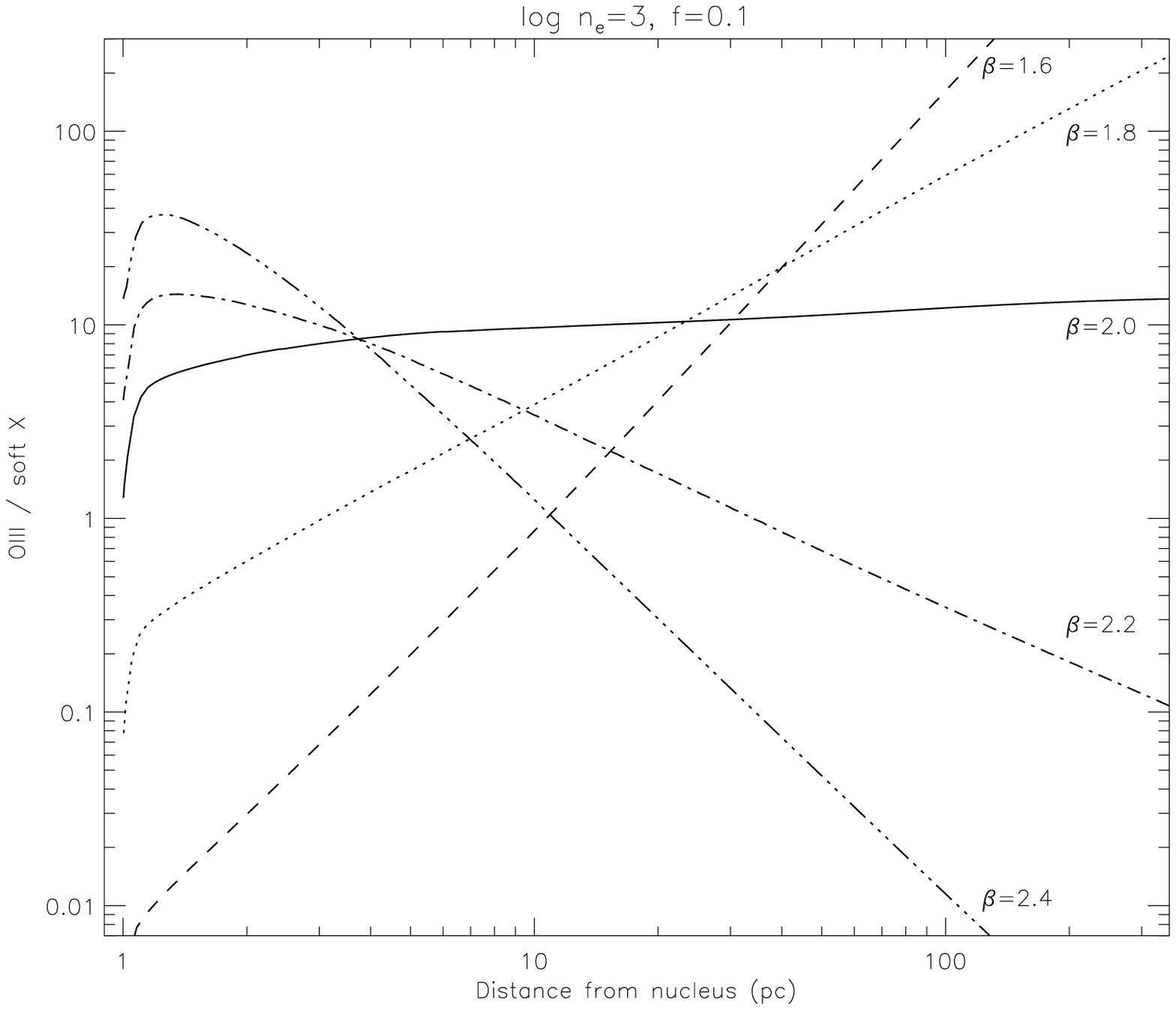, width=7.5cm}
\end{center}
\caption{\label{cone_ratio_r}Left: Each symbol represents one solution in our grid of \textsc{Cloudy} models that satisfies the condition of total [{O\,\textsc{iii}}] to soft X-ray flux ratio in the range observed in our sample, plotted in a three-parameter space $U$, $n_e$ and $\beta$, i.e. the ionization parameter and the density at the beginning of the cone of gas (1 pc), and the index of the density powerlaw, represented by different symbols (\textit{triangles}: $\beta=1.6$, \textit{diamonds}: $\beta=1.8$, \textit{stars}: $\beta=2.0$, \textit{squares}: $\beta=2.4$). The horizontal lines determine the limit corresponding to a total column density of $10^{20}$ cm$^{-2}$ for each index: solutions below this limit are not plotted. Solutions with $\beta=2.2$ are not plotted for clarity reasons. The net effect of the filling factor $f$ is to shift the density values: the two y axes refer to $f=0.1$ and $f=0.01$ (see text for details). Right: The [{O\,\textsc{iii}}] to soft X-ray ratio plotted as a function of the radius of the gas, for different values of $\beta$, when $\log{n_e}=3$ and $f=0.1$. The corresponding values of $\log{U}$ for these solutions are (from top to bottom): -0.85, -1.15, -1.4, -1.5, -1.6.}
\end{figure*}

The ratio between the [{O\,\textsc{iii}}] $\lambda5007$ line and the soft X-ray emission (defined as the total flux of the K$\alpha$ and K$\beta$ emission lines from N, O, Mg, S, Si in the range 0.5-2.0 keV) was calculated for each set of parameters. In the left panel of Fig. \ref{cone_ratio_r}, each symbol represents a solution in the $U$ versus $n_e$ plane, where this ratio has a value within 2.8 and 11, as observed in our sample. Different symbols correspond to different values of $\beta$. The net effect of changing the filling factor is simply a shift of the solutions along the density axis, by a factor equivalent to the variation in $f$, thus reproducing the same total column density for each set of three parameters constituting a `good' solution. The solutions occupy well-defined regions in the $n_e-U$ diagram, with those with lower $\beta$ being at larger values of ionization parameters.

The reason for this behaviour becomes clear inspecting the right panel of Fig. \ref{cone_ratio_r}, where the [{O\,\textsc{iii}}] to soft X-ray flux ratio is plotted as a function of the radius of the gas. Since $U\propto n_e^{-1} r^{-2}$, all density laws with $\beta<2$ produce a gas with a ionization parameter decreasing along with the distance. In these cases ($\beta=1.6$ and 1.8 in Fig. \ref{cone_ratio_r}), most of the soft X-ray emission is produced in the inner radii of the cone, while the bulk of the [{O\,\textsc{iii}}] emission is produced farther away, where the gas is less ionized. If $\beta=2$, the ionization parameter remains fairly uniform up to large radii, so that the total observed ratio between [{O\,\textsc{iii}}] and soft X-ray is constant with radius. Finally, if $\beta>2$, the ionization parameter increases with radius, so most of the soft X-rays are actually produced at larger radii, while the [{O\,\textsc{iii}}] emission line is mostly concentrated around the nucleus (cases $\beta=2.2$ and 2.4 in Fig. \ref{cone_ratio_r}). This is the reason why solutions with $\beta<1.6$ and $\beta>2.4$ are not plotted in Fig. \ref{cone_ratio_r}, even if these exist, satisfying the overall [{O\,\textsc{iii}}] to soft X-ray flux ratio. Their radial behaviour is radically different from what seen in Fig. \ref{xray2oiiimaps} and \ref{xray2oiiimaps_2}, where both the [{O\,\textsc{iii}}] and the soft X-ray emission are produced up to large radii. In particular, it is worth noting that constant-density solutions are totally unacceptable. A detailed analysis, as presented in Bianchi et al. (2005, submitted), suggests that the [{O\,\textsc{iii}}] to soft X-ray flux ratio observed in the sources of our sample remains fairly constant up to large radii, thus requiring that the density decreases  roughly like $r^{-2}$, similarly to what often found for the Narrow Line Region.

\begin{flushleft}
\bibliographystyle{aa}
\bibliography{sbs}
\end{flushleft}

\end{document}